\documentclass{svproc}

\usepackage{graphicx}
\usepackage{multirow}
\usepackage{amsmath,amssymb,amsfonts}
\usepackage{mathrsfs}
\usepackage[title]{appendix}
\usepackage{xcolor}
\usepackage{textcomp}
\usepackage{manyfoot}
\usepackage{booktabs}
\usepackage{algorithm}
\usepackage{algorithmicx}
\usepackage{algpseudocode}
\usepackage{listings}
\usepackage{cite}
\usepackage{hyperref}

\usepackage{url}

\begin{document}
\mainmatter 

\title{Bridging Semantics \& Structure for Software Vulnerability Detection using Hybrid Network Models}

\titlerunning{Bridging Semantics \& Structure for Vuln. Detection}  

\author{Jugal Gajjar\inst{1} \and Kaustik Ranaware\inst{1} \and Kamalasankari Subramaniakuppusamy\inst{1}}

\authorrunning{Jugal Gajjar et al.} 

\institute{$^1$The George Washington University, Washington D.C, USA\\
\email{jugal.gajjar@gwu.edu, k.ranaware@gwu.edu, and kamalasankaris@gwu.edu}\
}

\maketitle 

\begin{abstract}

Software vulnerabilities remain a persistent risk, yet static and dynamic analyses often overlook structural dependencies that shape insecure behaviors. Viewing programs as heterogeneous graphs, we capture control- and data-flow relations as complex interaction networks. Our hybrid framework combines these graph representations with light-weight ($<$4B) local LLMs, uniting topological features with semantic reasoning while avoiding the cost and privacy concerns of large cloud models. Evaluated on Java vulnerability detection (binary classification), our method achieves 93.57\% accuracy—an 8.36\% gain over Graph Attention Network-based embeddings and 17.81\% over pretrained LLM baselines such as Qwen2.5 Coder 3B. Beyond accuracy, the approach extracts salient subgraphs and generates natural language explanations, improving interpretability for developers. These results pave the way for scalable, explainable, and locally deployable tools that can shift vulnerability analysis from purely syntactic checks to deeper structural and semantic insights, facilitating broader adoption in real-world secure software development.


\end{abstract}

\section{Introduction}
\label{sec:intro}

Software vulnerabilities remain a critical threat to modern computing systems, from enterprise platforms to critical infrastructure. Real-world incidents such as Log4Shell \cite{log4shell} and the SolarWinds \cite{solarwinds} compromise highlight how flaws can propagate across interconnected components with devastating impact. Large software projects can be naturally modeled as complex networks, where nodes correspond to program elements (e.g., functions, statements) and edges capture control- or data-flow dependencies. These networks exhibit structural properties such as modularity, degree distributions, and path dependencies that shape how vulnerabilities emerge and spread \cite{hao2020}. Capturing such dynamics is key to detecting flaws often missed by traditional analyses.

Conventional representations such as control-flow graphs (CFGs) or abstract syntax trees (ASTs) highlight local structure but fail to capture global connectivity or semantic intent \cite{harzevili2023, alashjaee2019}. In contrast, large language models (LLMs) excel at semantic reasoning over token sequences but overlook program topology, limiting their ability to model complex multi-element interactions \cite{tie2024}. Graph embeddings capture connectivity yet struggle to encode execution semantics \cite{yu2023, alon2020}. The challenge, therefore, is clear: \emph{how can structural properties of program networks and semantic reasoning of LLMs be fused for effective vulnerability detection?}

To the best of our knowledge, this is the first framework that unifies graph-based embeddings of Java programs with lightweight ($<$4B) local LLMs. Specifically, we (1) design a two-way gating mechanism that adaptively aligns structural and semantic spaces, (2) apply contrastive learning with InfoNCE objectives \cite{infonce} to enforce cross-view consistency, (3) incorporate graph Laplacian regularization \cite{ando2006} to promote smoothness across directed CFGs, and (4) deliver interpretability through top-K saliency subgraphs with natural language explanations.  

By foregrounding interpretability alongside accuracy, our approach ensures developers gain actionable insights rather than opaque predictions. Ultimately, we demonstrate that fusing network-theoretic and semantic perspectives yields state-of-the-art vulnerability detection with transparency, scalability, and deployability in privacy-preserving settings.

\section{Related Work}
\label{sec:rel_work}

Traditional static analysis techniques detect vulnerabilities through fixed rulesets, offering precision for certain flaw classes but suffering from high false positives and limited ability to capture runtime-dependent issues \cite{kaur2020, charoenwet2024}. These approaches emphasize local code structure without modeling the broader network of interactions among program elements \cite{harzevili2023}. Dynamic analysis, by monitoring execution, exposes runtime errors such as leaks or deadlocks but struggles with incomplete path coverage and high testing overheads \cite{alashjaee2019}.

ML and DL methods extend these approaches by learning vulnerability patterns from large datasets. Sequence-based methods such as VulDeePecker \cite{zou2019} and SySeVR \cite{li2021} capture local ordering of tokens, while MalCodeAI \cite{gajjar2025} integrates LLM-based detectors with explainability. However, they underutilize the complex network structure of software, where vulnerabilities often arise from cycles, hubs, and long-range dependencies—phenomena well-studied in network science but rarely modeled in vulnerability detection.

LLMs trained on source code exhibit strong semantic reasoning \cite{yu2025}, yet their limited context windows hinder capturing higher-order connectivity across functions and modules. Graph-based models such as Devign \cite{zhou2019} leverage ASTs or CFGs to encode dependencies, but often reduce networks to local neighborhoods, missing global structural features like motifs or community-level interactions that may reveal systemic weaknesses. This gap mirrors broader findings in complex network research, where global topology and local patterns jointly explain emergent behaviors.

Recent GNN advances for vulnerability detection show promise but face key limitations. VulGraB \cite{vulgrab} uses bi-directional gated GNNs but focuses on local neighborhoods, missing global patterns. Yang et al. \cite{yang2024} propose tensor-based gated GNNs that struggle with long-range dependencies, while RGAN \cite{rgan} uses residual attention networks but misses higher-order motifs. These graph approaches excel at structural modeling but fail to capture code semantics essential for subtle vulnerability patterns, treating code as abstract nodes without semantic richness. Additionally, most work targets low-level languages (C/C++, Assembly) using benchmarks like Devign \cite{zhou2019}, Juliet \cite{boland2012}, and MegaVul \cite{ni2024}, while Java remains underexplored despite widespread production deployment and unique runtime characteristics.

Our work addresses these limitations by integrating network-theoretic and ML perspectives: we model programs as directed graphs to capture higher-order execution dependencies and fuse them with lightweight local LLM embeddings for semantic reasoning. To the best of our knowledge, this is the first framework to apply such a hybrid approach to Java, enabling scalable, privacy-preserving, and interpretable vulnerability detection.

Table~\ref{tab:rel_work_comparison} summarizes how different approaches address vulnerability detection.

\begin{table}[tb]
\centering
\caption{Comparison of approaches for software vulnerability detection.}
\label{tab:rel_work_comparison}
\begin{tabular}{@{}lccccc@{}}
\toprule
\textbf{Method} & \textbf{Vul. Detect.} & \textbf{Code Behav.} & \textbf{Code Cover.} & \textbf{Structural} & \textbf{Semantics} \\
\midrule
Static Anal. & $\checkmark$ & $\times$ & $\times$ & $\checkmark$ & $\times$ \\
Dynamic Anal. & $\checkmark$ & $\checkmark$ & $\times$ & $\times$ & $\times$ \\
ML/DL & $\checkmark$ & $\checkmark$ & $\times$ & $\times$ & $\checkmark$ \\
LLMs & $\checkmark$ & $\checkmark$ & $\times$ & $\times$ & $\checkmark$ \\
Graph Models & $\checkmark$ & $\times$ & $\checkmark$ & $\checkmark$ & $\times$ \\
\textbf{Our Work} & $\checkmark$ & $\checkmark$ & $\checkmark$ & $\checkmark$ & $\checkmark$ \\
\bottomrule
\end{tabular}
\end{table}

\section{Methodology}
\label{sec:methodology}

This section describes the proposed hybrid framework that fuses graph-structured program representations with LLM-derived semantic embeddings for vulnerability detection. We first present the code-as-network representation and embedding extraction pipeline, then formalize the projection and fusion modules (with emphasis on our proposed two-way gating fusion), present the composite training objective (classification + contrastive alignment \cite{infonce} + Laplacian regularization \cite{ando2006}), and finally describe the explanation mechanisms used at inference.

\subsection{Code as a Graph}
We represent each program unit (function / file) as a \emph{control-flow graph} (CFG) \( \mathcal{G} = (V,E) \) where nodes \(v\in V\) correspond to program statements or basic blocks and directed edges \( (u\to v)\in E\) represent control- or data-flow transitions. The graph is encoded by its adjacency matrix \(\mathbf{A}\in\{0,1\}^{|V|\times|V|}\) and degree matrix \(\mathbf{D}=\mathrm{diag}(d_1,\dots,d_{|V|})\). We consider both unnormalized Laplacian \(\mathbf{L}=\mathbf{D}-\mathbf{A}\) and the normalized variant \(\mathbf{L}_{\mathrm{sym}}=\mathbf{I}-\mathbf{D}^{-1/2}\mathbf{A}\mathbf{D}^{-1/2}\) when appropriate.

Modeling software as a complex network enables the use of graph-signal priors (e.g., Laplacian smoothness) and network-theoretic insights to regularise learning and improve interpretability.

\subsection{Embedding Extraction}
For each sample \(i\) we extract two complementary embeddings:

\subsubsection{Graph embeddings \(h_{G,i}\).} Node features are constructed from node text (e.g., statement tokens or code snippets). We extract graph-level embeddings using four complementary encoders: GraphSAGE \cite{graphsage} for neighborhood sampling and aggregation, GCN \cite{gcn} for message passing, GAT \cite{gat} as a graph attention network that learns edge attention coefficients, and Node2Vec \cite{node2vec} for random-walk based structural embeddings. Each encoder produces node embeddings which are pooled (global mean pooling) to obtain a graph-level vector \(h_{G,i}\in\mathbb{R}^{d_G}\). In this work we fix \(d_G=128\) for all graph pipelines to ensure a fair comparison between encoders.

\subsubsection{LLM embeddings \(h_{L,i}\).} For each code sample we obtain a contextual embedding \(h_{L,i}\in\mathbb{R}^{d_L}\) from a local code-Large-Language-Model (LLM). To preserve privacy and enable execution on modest hardware we restrict to lightweight models (parameter count $<$4B). Models considered include Qwen2.5 Coder 3B \cite{qwen25}, DeepSeek Coder 1.3B \cite{deepseekcoder}, Stable Code 3B \cite{stablecoder}, and Phi 3.5 Mini \cite{phi35}. LLM embeddings encode high-order semantics (name usage, API semantics, docstrings, idioms) that structural encoders cannot directly capture.

\subsubsection{Normalization.} For numerical stability we L2-normalize embeddings before projection:
\begin{align}
\tilde{h}_{G,i} \leftarrow \frac{h_{G,i}}{\|h_{G,i}\|_2 + \epsilon}, \qquad
\tilde{h}_{L,i} \leftarrow \frac{h_{L,i}}{\|h_{L,i}\|_2 + \epsilon},
\end{align}
where \(\epsilon\) is a small constant.

\subsection{Projection and alignment}
Graph and LLM embeddings typically live in different dimensionalities (\(d_G\) vs \(d_L\)). We therefore learn linear projection heads that map both modalities into a common latent space \(\mathbb{R}^{d'}\):
\begin{align}
\hat{g}_i &= \mathrm{LayerNorm}\big(\phi(W_g \tilde{h}_{G,i} + b_g)\big), \qquad W_g\in\mathbb{R}^{d'\times d_G},\\
\hat{l}_i &= \mathrm{LayerNorm}\big(\phi(W_l \tilde{h}_{L,i} + b_l)\big), \qquad W_l\in\mathbb{R}^{d'\times d_L},
\end{align}
where \(\phi(\cdot)\) is a nonlinearity (GELU/ReLU) and LayerNorm stabilizes distributions. In our implementation \(d'=128\) by default (configurable).

Projecting before fusion reduces the risk that one modality dominates due to scale or dimension mismatch, and creates a compact, shared latent space for contrastive alignment.

\subsection{Fusion architectures}
We compare three fusion families; the two-way gating is proposed as our primary fusion module.

\subsubsection{(A) Early concatenation (baseline).}
Concatenate projected vectors and use an MLP classifier:
\begin{align}
u_i = [\hat{g}_i \,\|\, \hat{l}_i] \in \mathbb{R}^{2d'},\qquad
s_i = f_{\mathrm{MLP}}(u_i),\qquad \hat{y}_i=\sigma(s_i).
\end{align}
This is simple and effective but does not adaptively re-weight modalities per-example.

\subsubsection{(B) \textbf{Two-way Gating} (proposed).}
We propose a two-way gating mechanism that \emph{adaptively} combines structural and semantic signals at sample level, and — crucially — yields an interpretable scalar weight indicating modality importance for the decision.

Concretely, let \([\hat{g}_i\|\hat{l}_i]\) denote concatenation. We compute scalar gating scores via small scoring networks and convert to a two-way soft attention:
\begin{align}
e_{g,i} &= v^\top \tanh\!\big( W_e [\hat{g}_i \,\|\, \hat{l}_i] + b_e\big),\\
e_{l,i} &= v^\top \tanh\!\big( W_e' [\hat{l}_i \,\|\, \hat{g}_i] + b_e'\big),\\
[a_{g,i},a_{l,i}] &= \mathrm{softmax}\big([e_{g,i}, e_{l,i}]\big),\\
\hat{h}_i &= a_{g,i}\,\hat{g}_i + a_{l,i}\,\hat{l}_i.
\end{align}
Final classification proceeds from \(\hat{h}_i\) via an MLP.

\textbf{Why two-way gating?} Unlike a fixed concatenation or a global linear combination, two-way gating learns \emph{per-sample} preferences: in some functions structural layout (e.g., complex branching) is decisive and \(a_{g}\) will dominate; in others semantic hints (sensitive API call patterns) are decisive and \(a_{l}\) will dominate. The architecture is compact, interpretable (the scalar weights are human-readable), and empirically reduces error and improves calibration, particularly when modalities disagree.

\subsubsection{(C) QKV Cross-attention.}
We also evaluate a cross-attention fusion where one modality queries the other:
\begin{align}
Q_i &= W_q \hat{l}_i,\quad K_j = W_k \hat{g}_j,\quad V_j = W_v \hat{g}_j,\\
\alpha_{ij} &= \mathrm{softmax}_j\Big( \frac{Q_i K_j^\top}{\sqrt{d_k}} \Big),\qquad
F_i = \sum_j \alpha_{ij} V_j,
\end{align}
then optionally combine \(F_i\) with \(\hat{l}_i\). Cross-attention allows the LLM embedding to selectively attend to structural motifs, but it increases parameter count and may overfit on small datasets (see the Transformer architecture in \cite{vaswani2017}).

\subsection{Classification head}

Across all fusion modules, we use a simple classification mapping that concatenates graph and language representations before applying a linear layer with sigmoid activation. The training objective operates on the logit \(s_i = W_c u_i + b_c\).

\subsection{Training objectives}
We train the fusion model with a composite objective that (i) enforces correct classification, (ii) aligns graph and LLM embeddings, and (iii) imposes smoothness across the program graph.

\subsubsection{(1) Binary classification loss (BCE).}
For \(N\) training samples with labels \(y_i\in\{0,1\}\):
\begin{equation}
\mathcal{L}_{\mathrm{cls}} = -\frac{1}{N}\sum_{i=1}^N\Big[y_i\log\hat{y}_i + (1-y_i)\log(1-\hat{y}_i)\Big],
\end{equation}
which directly optimizes the downstream vulnerability prediction accuracy.

\subsubsection{(2) Contrastive alignment (InfoNCE).}
To encourage the projected graph and LLM embeddings of the \emph{same} sample to be similar while pushing apart others, we use an in-batch InfoNCE loss \cite{infonce}. Let \(\hat{g}_i,\hat{l}_i\) be projected and L2-normalized. Define similarity \(\mathrm{sim}(x,y)=x^\top y\) (cosine after normalization). Then:
\begin{equation}
\mathcal{L}_{\mathrm{InfoNCE}} = -\frac{1}{N}\sum_{i=1}^N
\log\frac{\exp\big(\mathrm{sim}(\hat{g}_i,\hat{l}_i)/\tau\big)}
{\sum_{j=1}^N\exp\big(\mathrm{sim}(\hat{g}_i,\hat{l}_j)/\tau\big)},
\end{equation}
with temperature \(\tau>0\). This term (i) stabilizes cross-modal training, (ii) prevents the fusion head from ignoring one modality, and (iii) yields a shared semantic-structural manifold that benefits prototype retrieval and nearest-neighbour explainability.

\subsubsection{(3) Graph Laplacian regularizer.}
To encourage neighboring nodes in each CFG to have smoothly-varying embeddings (i.e., to respect local structural coherence), we apply a Laplacian regularizer \cite{ando2006}. Let \(\mathbf{H}\in\mathbb{R}^{|V|\times d'}\) be the matrix of node embeddings for a graph (in the projection space). The Laplacian penalty is:
\begin{equation}
\mathcal{L}_{\mathrm{Lap}} = \frac{1}{M}\sum_{m=1}^M \mathrm{Tr}\big(\mathbf{H}_m^\top \mathbf{L}_m \mathbf{H}_m\big)
= \frac{1}{M}\sum_{m=1}^M \sum_{(u,v)\in E_m} \|h_u - h_v\|_2^2,
\end{equation}
averaged over the \(M\) graphs (batched). This term enforces structural smoothness and is especially meaningful for directed CFGs where neighboring blocks should preserve execution-consistent signals.

\subsubsection{(4) Final objective.}
We combine the components with scalar weights:
\begin{equation}
\boxed{\; \mathcal{L} = \mathcal{L}_{\mathrm{cls}} \;+\; \lambda_{\mathrm{nce}}\,\mathcal{L}_{\mathrm{InfoNCE}} \;+\; \lambda_{\mathrm{lap}}\,\mathcal{L}_{\mathrm{Lap}} \; }
\end{equation}
Hyperparameters \(\lambda_{\mathrm{nce}}\) and \(\lambda_{\mathrm{lap}}\) control the strength of alignment and structural smoothness; values are chosen by validation.

\subsection{Interpretability and one-line explanations}
Interpretability is essential for vulnerability triage. Our pipeline produces three complementary artifacts for each prediction:
\begin{enumerate}
  \item \textbf{Gating weights} \((a_{g,i},a_{l,i})\) from the two-way fusion — a per-sample scalar explanation indicating whether structure or semantics drove the decision.
  \item \textbf{Gradient-based saliency} on projected node features (Integrated Gradients or input-gradients) to highlight the top-\(K\) most influential CFG nodes.
  \item \textbf{One-line LLM justification.} We generate a concise, single-sentence justification using the local LLM conditioned on the code, predicted label, and the top-\(K\) salient nodes. Strict prompt template (deterministic decoding) ensures consistent, short explanations:
\begin{quote}
\texttt{One sentence: main reason this code is \{"vulnerable" if cls == 1 else "safe"\}. Provide only one sentence.\\ Code: <CODE>\\ Top nodes: <node\_labels>\\ One-sentence explanation:}
\end{quote}
This LLM-written explanation improves human interpretability while remaining grounded in the model's own internal saliency signals (we feed the salient node labels as context).
\end{enumerate}

\begin{figure}[tb]
\centering
\includegraphics[width=\linewidth, height=2.15in]{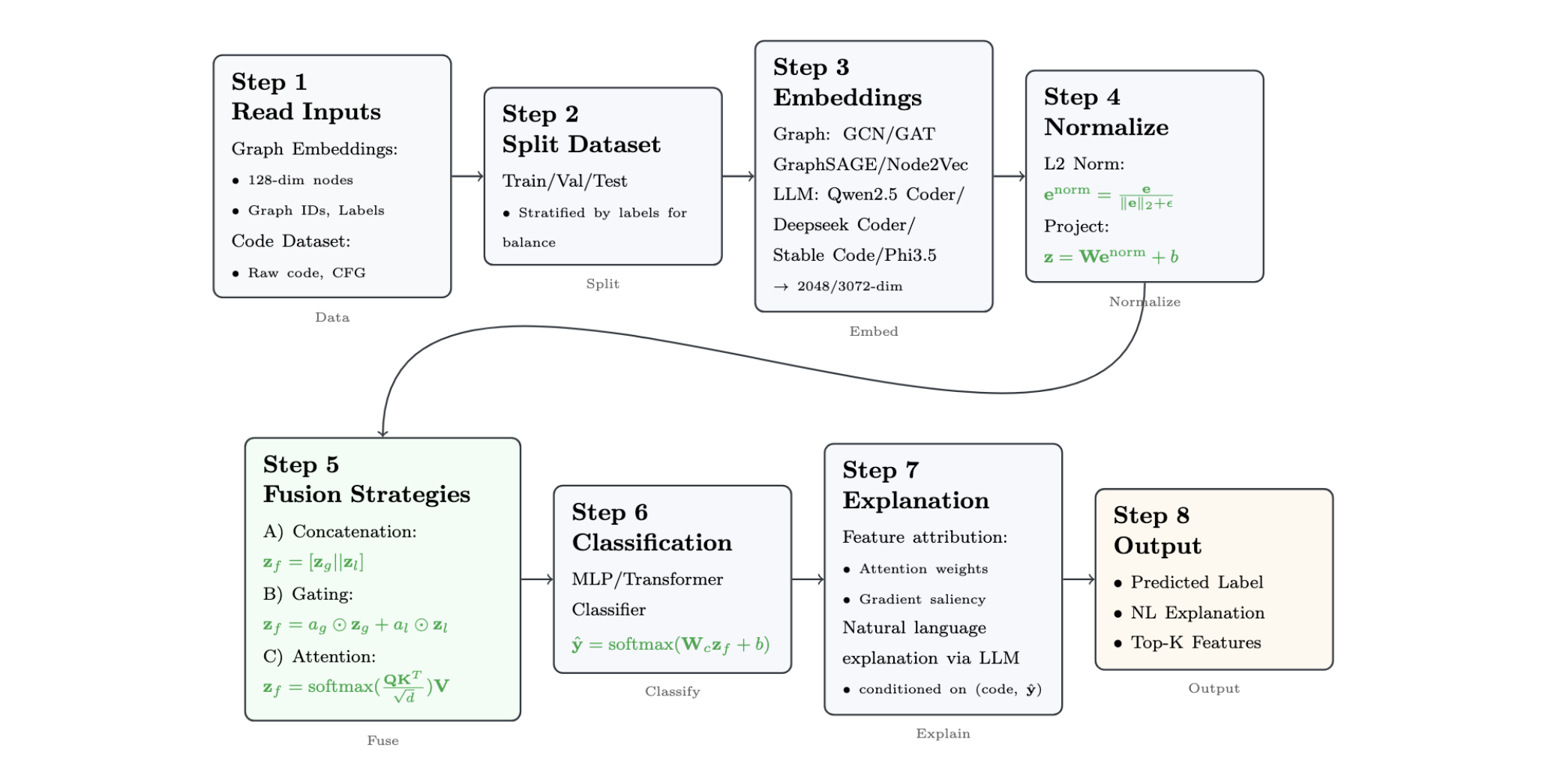}
\caption{Multimodal fusion pipeline for code classification with explainability. The pipeline processes graph embeddings and code text through normalization, fusion, classification, and explanation generation stages.}
\label{fig:pipeline}
\end{figure}

\section{Experiments and Results}
\label{sec:exp_and_res}

This section presents our empirical validation, moving from data collection to interpretability. We start with dataset preparation, then isolate the effects of structural and semantic representations, before evaluating their fusion and robustness through ablations. Finally, we show that the model is both accurate and explainable. These results affirm our thesis: \emph{graphs encode indispensable structure, semantics enrich representation, and fusion yields the strongest and most interpretable models}.

\subsection{Dataset}
We curated a large Java dataset from Vul4J \cite{vul4j}, Kaggle CVE Fixes \cite{cvefixes}, and Kaggle Vulnerability Fixes \cite{kagglevulnfixes}, collecting 36,425 code samples. Due to heavy vulnerability bias, we created 6,000 synthetic safe files by systematically stripping vulnerable patterns from real-world snippets (e.g., adding input validation, fixing resource handling), ensuring compilable, functionally safe code. Other works using equivalent corpora either employ similar practices or accept the imbalance, but we found balancing necessary to prevent trivial decision boundaries that would otherwise inflate performance. After removing duplicates and erroneous codes, the final dataset contained 35,610 CFG-parsable files (6,234 safe, 29,376 vulnerable). The dataset is publicly available at \href{https://doi.org/10.5281/zenodo.17259519}{zenodo.org/records/17259519}. Control Flow Graphs were generated using PROGEX \cite{progex}, averaging 20.6 nodes and 20.3 edges per graph, enabling simultaneous analysis of structural complexity and semantic intent via pretrained LLM embeddings. Code for data preprocessing, model training, and evaluation is available on our GitHub repository: \href{https://github.com/JugalGajjar/VulnGraph}{github.com/JugalGajjar/VulnGraph}.

\subsection{Structural Information via Graphs}
We first evaluate purely structural models: GCN \cite{gcn}, GAT \cite{gat}, GraphSAGE \cite{graphsage}, and Node2Vec \cite{node2vec}. These architectures capture local and global structural motifs of CFGs and serve as our backbone encoders. Table~\ref{tab:graph_results} reports test accuracy. Across all models, graph-based methods outperform shallow baselines, demonstrating the importance of structural signals in vulnerability detection.

\begin{table}[tb]
\centering
\caption{Performance of graph networks before and after adding semantics.}
\label{tab:graph_results}
\begin{tabular}{@{}lccc@{}}
\toprule
\textbf{Architecture} & \textbf{Acc. (Graph Only)} & \textbf{Acc. (+Semantics)} & \textbf{$\Delta$} \\
\midrule
GCN & 73.08 & 81.12 & 8.04 \\
GAT & 85.21 & 90.56 & 5.35 \\
GraphSAGE & 80.77 & 87.55 & 6.78 \\
Node2Vec & 72.14 & 84.50 & 12.36 \\
\bottomrule
\end{tabular}
\end{table}

\subsection{Semantic Augmentation}
To assess the contribution of semantic knowledge, we augment each graph encoder with GraphCodeBERT \cite{graphcodebert} embeddings. Table~\ref{tab:graph_results} shows that across all architectures, semantic enrichment improves performance, confirming that structural and semantic cues are complementary. The performance gain is especially pronounced for shallow encoders like Node2Vec \cite{node2vec}, which benefit strongly from added semantic context.

\begin{figure}[tb]
\centering
\includegraphics[width=0.8\linewidth, height=2.3in]{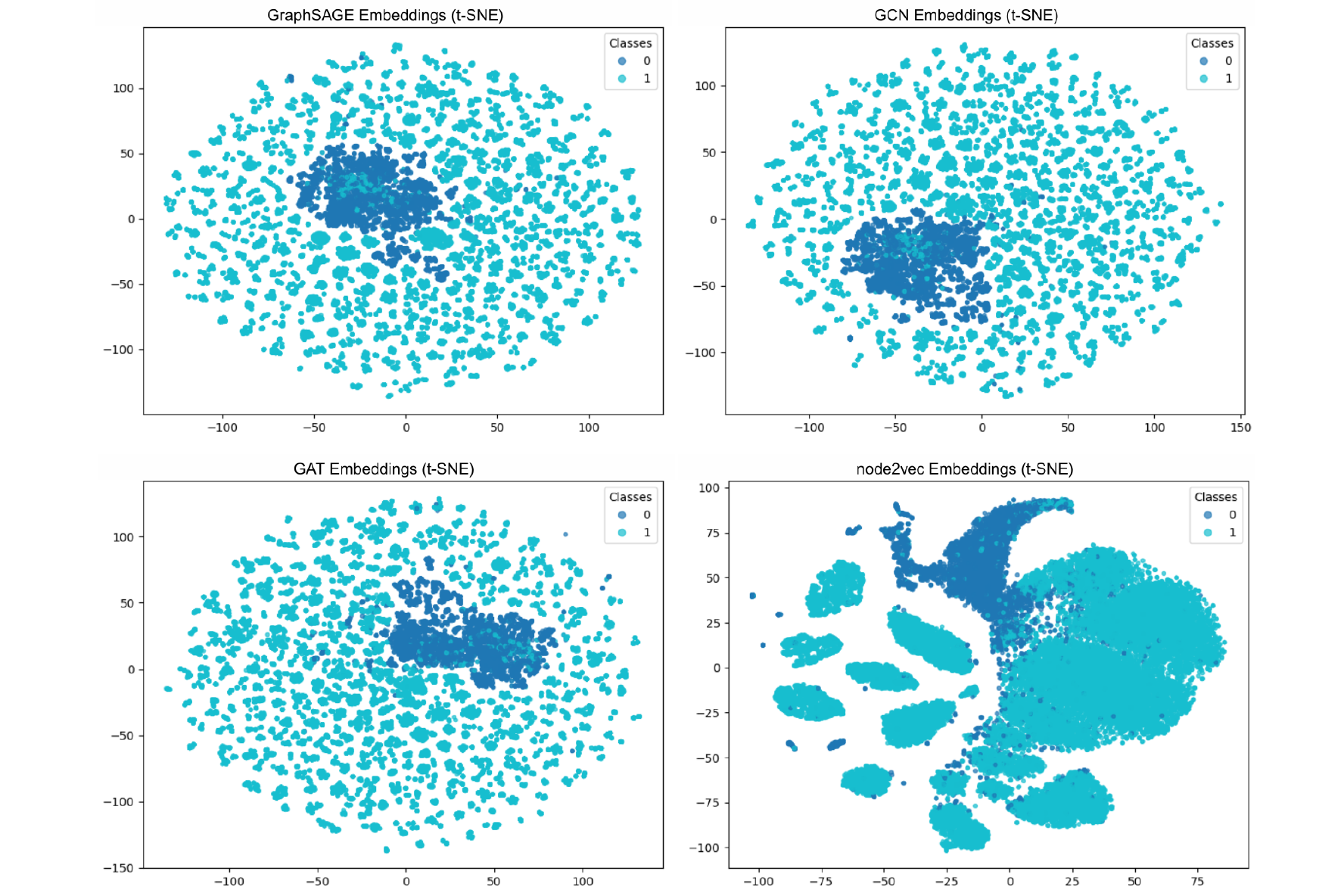}
\caption{t-SNE visualization of graph embeddings for four architectures with GraphCodeBERT semantics.}
\label{fig:graph_visualizations}
\end{figure}

\subsection{Zero-Shot LLM Classification}
We further probe pretrained LLMs directly via zero-shot prompting, asking them to classify Java code as safe or vulnerable without task-specific training. While performance is lower than trained graph encoders, these results establish a baseline for purely semantic, non-structural reasoning. Refer Table~\ref{tab:llm_zeroshot} for results.

\begin{table}[tb]
\centering
\caption{Zero-shot classification by pretrained LLMs.}
\label{tab:llm_zeroshot}
\begin{tabular}{@{}lccc@{}}
\toprule
\textbf{LLM} & \textbf{Accuracy (\%)} & \textbf{Std. Dev. over Runs} & \textbf{Notes} \\
\midrule
Deepseek Coder 1.3B & 62.43 & 2.76 & Limited reasoning \\
Qwen2.5 Coder 3B & 75.76 & 1.62 & Strong baseline \\
Stable Code 3B & 51.01 & 3.43 & Weak generalization \\
Phi3.5 Mini 3.8B & 68.29 & 4.07 & Moderate performance \\
\bottomrule
\end{tabular}
\end{table}

\subsection{Graph-Semantic Fusion}
We then fuse graph and semantic embeddings using three strategies: early concatenation, two-way gating, and cross-attention. Experiments were run across all combinations of LLMs and graph encoders. Results (Table~\ref{tab:fusion}) show that fusion consistently outperforms either modality alone, with two-way gating achieving the best trade-off between performance and interpretability. 

\begin{table}[tb]
\centering
\caption{Fusion results across graph encoders and LLMs.}
\label{tab:fusion}
\begin{tabular}{lcc}
\toprule
\textbf{Fusion Strategy} & \textbf{Accuracy (\%)} & \textbf{Notes} \\
\midrule
Concatenation & 91.14 & Simple baseline \\
Two-way Gating & 93.57 & Best overall \\
Cross-Attention & 92.07 & Strong but unstable \\
\bottomrule
\end{tabular}
\end{table}

\subsection{Ablation Study}
To understand the contributions of individual components in our best performing system, we perform ablations by (i) removing graph features, (ii) removing LLM semantic features, (iii) removing InfoNCE contrastive loss \cite{infonce}, and (iv) removing Laplacian regularization \cite{ando2006}. Table~\ref{tab:ablations} shows that graphs are the strongest individual factor, while contrastive and Laplacian terms further improve robustness and generalization.  

\begin{table}[tb]
\centering
\caption{Ablation study on fusion model.}
\label{tab:ablations}
\begin{tabular}{@{}lcc@{}}
\toprule
Model Variant & Accuracy (\%) & Drop By \\
\midrule
Full (Graph + Semantics + Losses) & 93.57 & – \\
No Graphs & 78.93 & 14.64 \\
No Semantics & 85.21 & 8.36 \\
No InfoNCE & 88.28 & 5.29 \\
No Laplacian & 86.72 & 6.85 \\
\bottomrule
\end{tabular}
\end{table}

\subsection{Interpreting Fusion Models}
\begin{figure}[tb]
\centering
\includegraphics[width=0.91\linewidth]{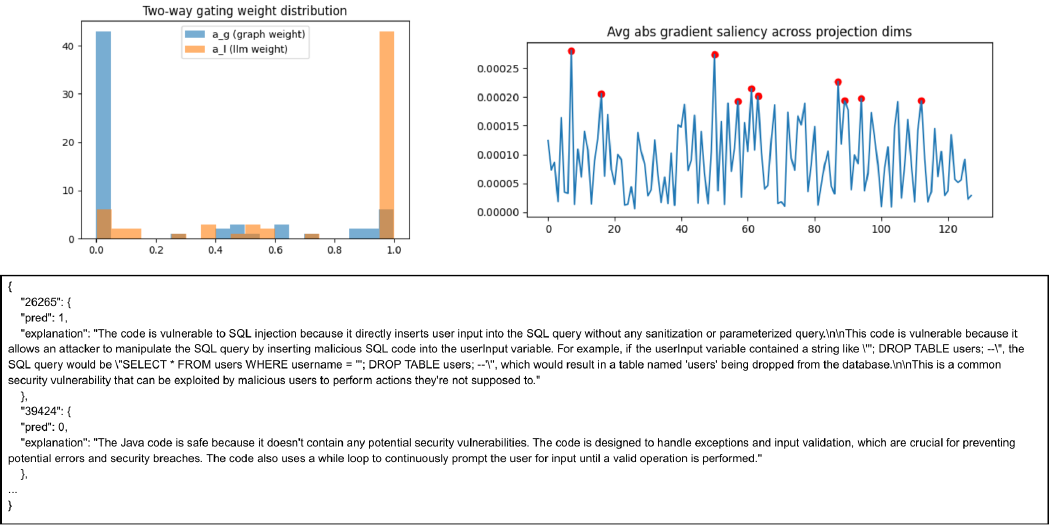}
\caption{Top-left: Two-way gating weight distribution. Top-right: Gradient saliency across projection dimensions. Bottom: Example LLM-generated justifications.}
\label{fig:explainability}
\end{figure}

Finally, we investigate interpretability (Figure~\ref{fig:explainability}). We use gradient-based saliency to identify the most influential structural features, and gating weight visualization to show modality contributions. Explanations are also generated by the same LLM (with output capped at 256 tokens), providing human-readable rationales. While raw saliency values across projection dimensions are hard to interpret, we aggregate them into node-level importance for clearer signals. Gating weights are likewise contextualized by linking high-weight cases to supporting semantic or structural evidence in the LLM explanations.

\section{Discussion and Future Work}
\label{sec:dis_and_fut_work}

Our experiments show that combining Laplacian smoothness with structural-semantic fusion yields significant improvements in vulnerability detection. Laplacian regularization enforces local consistency in embeddings, reducing overfitting and enabling the model to capture distributed structural dependencies across a program. At the same time, fusing graph encoders with large language models unites control- and data-flow relationships with semantic and contextual cues, more faithfully reflecting the complex, network-like nature of software. This integration explains the substantial performance gains observed in our best-performing strategies.

Despite these advances, several limitations remain. Directed program graphs lack symmetry, complicating the use of spectral techniques that assume undirected structure. Scaling to ultra-large, industrial codebases will also require efficiency-focused optimizations. Future work should explore richer graph representations, including multiplex networks that integrate ASTs, CFGs, and dataflow; temporal graphs to track software evolution; and cross-language vulnerability detection. Moving beyond binary classification toward fine-grained, multiclass categorization of vulnerability types represents another important direction.

\section{Conclusion}
\label{sec:conclusion}

This work demonstrates that combining structural graph encoders with semantic language models, guided by contrastive learning and Laplacian smoothness, provides a powerful framework for software vulnerability detection. By unifying program structure and code semantics, our approach achieves superior performance while offering faithful representation of how vulnerabilities manifest in real-world systems. Our findings highlight the importance of treating software as both network and language—an insight that enables richer, more generalizable, and interpretable security analysis. While challenges remain in scalability and fine-grained classification, the methodological advances introduced here establish a solid foundation for future research in graph-based AI for secure software development.

\end{document}